\documentclass[12pt]{article}

\pdfoutput=1

\usepackage{graphics}
\usepackage{amssymb}
\usepackage{amsmath}
\usepackage{hyperref}
\usepackage{bbold}
\usepackage{epstopdf}
\usepackage{tikz}

\usepackage[style=authoryear-ibid,backend=biber,maxbibnames=6,maxcitenames=4,sorting=nyt,dashed=no,doi=false]{biblatex}
\DeclareNameAlias{sortname}{last-first}
\DeclareFieldFormat[article, inbook]{title}{#1}
\renewbibmacro{in:}
{%
\ifentrytype{article}{}
{%
\printtext{\bibstring{in}\intitlepunct}}}
\newcommand{\rf}[1]{(\ref{#1})}
\newcommand{\beq}{\begin{equation}}

\newcommand{\eeq}{\end{equation}}
\newcommand{\bea}{\begin{eqnarray}}
\newcommand{\eea}{\end{eqnarray}}
\newcommand{\e}{\mbox{e}}

\newcommand{\kp}{\kappa}

\newcommand{\la}{\langle}
\newcommand{\ra}{\rangle}
\newcommand{\plu}{\!+\!}
\newcommand{\mi}{\!-\!}
\newcommand{\bN}{\bar{N}}

\begin{document}

\begin{center}
${}$\\
\vspace{100pt}
{ \Large \bf Causal Dynamical Triangulations:
\\ \vspace{10pt} Gateway to Nonperturbative Quantum Gravity
}

\vspace{36pt}

{\sl J. Ambj\o rn}$\,^{a,b}$
and
{\sl R. Loll}$\,^{a}$

\vspace{18pt}

{\footnotesize
$^a$~Institute for Mathematics, Astrophysics and Particle Physics, Radboud University \\ 
Heyendaalseweg 135, 6525 AJ Nijmegen, The Netherlands.\\ 
{email: r.loll@science.ru.nl}\\

\vspace{6pt}

$^b$~The Niels Bohr Institute, Copenhagen University\\
Blegdamsvej 17, DK-2100 Copenhagen \O , Denmark.\\
{email: ambjorn@nbi.dk}\\

}

\end{center}

\vspace{20pt}

\begin{center}
{\bf Abstract}
\end{center}

\noindent 
A powerful strategy to treat quantum field theories beyond perturbation theory is by putting them on a lattice. However, the dynamical and symmetry structure of general relativity have for a long time stood in the way of a well-defined lattice formulation of quantum gravity. These issues are resolved by using Causal Dynamical Triangulations (CDT) to implement a nonperturbative, background-independent path integral for Lorentzian quantum gravity on dynamical lattices. We describe the essential ingredients of this formulation, and how it has allowed us to move away from formal considerations in quantum gravity to extracting quantitative results on the spectra of diffeomorphism-invariant quantum observables, describing physics near the Planck scale. Key results to date are the emergence of a de Sitter-like quantum universe and the discovery of an anomalous spectral dimension at short distances.

\vspace{1.0cm}

\newpage

\tableofcontents

\newpage

\section{Introduction}
\label{sec:intro}

To outsiders and practitioners alike, quantum gravity can at times appear as a subject whose aims, progression, relevance and perhaps very existence are forever elusive. Our search for a fundamental theory of quantum gravity is hampered by the extreme weakness of the gravitational interactions, which implies that large quantum effects are naturally associated with extreme energy scales. As a consequence, there are currently no experiments or observations 
to guide our theory-building, or help us discriminate between fruitful ideas and idle speculation.   

Another sense in which gravity is fundamentally different is the fact that its dynamical field is spacetime itself, in contrast with
the nongravitational fields of the standard model of particle physics, which propagate on a fixed, inert background spacetime.  
This has long given rise to speculations that the dynamical nature of spacetime in gravity may ultimately be incompatible with the basic tenets of
quantum field theory, and that the construction of a theory of quantum gravity therefore requires radically new ingredients 
and physical principles. 

By contrast, the theory and methodology presented here, subsumed under the name \textit{Causal Dynamical Triangulations} or \textit{CDT} for short, 
shows that there is a conceptually much simpler way to understand quantum gravity beyond perturbation theory. It uses purely quantum field-theoretic
concepts and a minimal set of ingredients and free parameters, but nevertheless produces robust results that are neither trivial nor obvious. 
Its quantum dynamics is defined in terms of the time-honoured \textit{gravitational path integral} or \textit{sum over histories}
\begin{equation}
Z=\! \int\! {\cal D} [g]\; \mathrm{e}^{\, iS^\mathrm{grav}[g]},
\label{path}
\end{equation}
where the superposition is taken over all spacetime geometries $[g]$, each with a complex weight depending on its gravitational action $S^\mathrm{grav}[g]$.
As usual in quantum field theory, the formal continuum path integral (\ref{path}) is ill-defined and highly divergent. It needs nontrivial input to transform it
from a mere statement of intent to a mathematically well-defined  description from which physical predictions can be derived

The first key point regarding this ansatz is that CDT provides a precise
prescription for regularizing (\ref{path}) in a manifestly diffeomorphism-invariant and background-independent way, which also 
incorporates the Lorentzian signature of spacetime. The spacetime configurations in the path integral are regularized in terms of dynamical lattices made from
small triangular Minkowskian building blocks, from which the formulation takes its name. To obtain a theory of quantum gravity one still needs to perform a
continuum limit, where the ultraviolet (short-distance) lattice regulator is removed and coupling constants are suitably renormalized.   

\begin{figure}[t]
\centerline{\scalebox{0.28}{\rotatebox{0}{\includegraphics{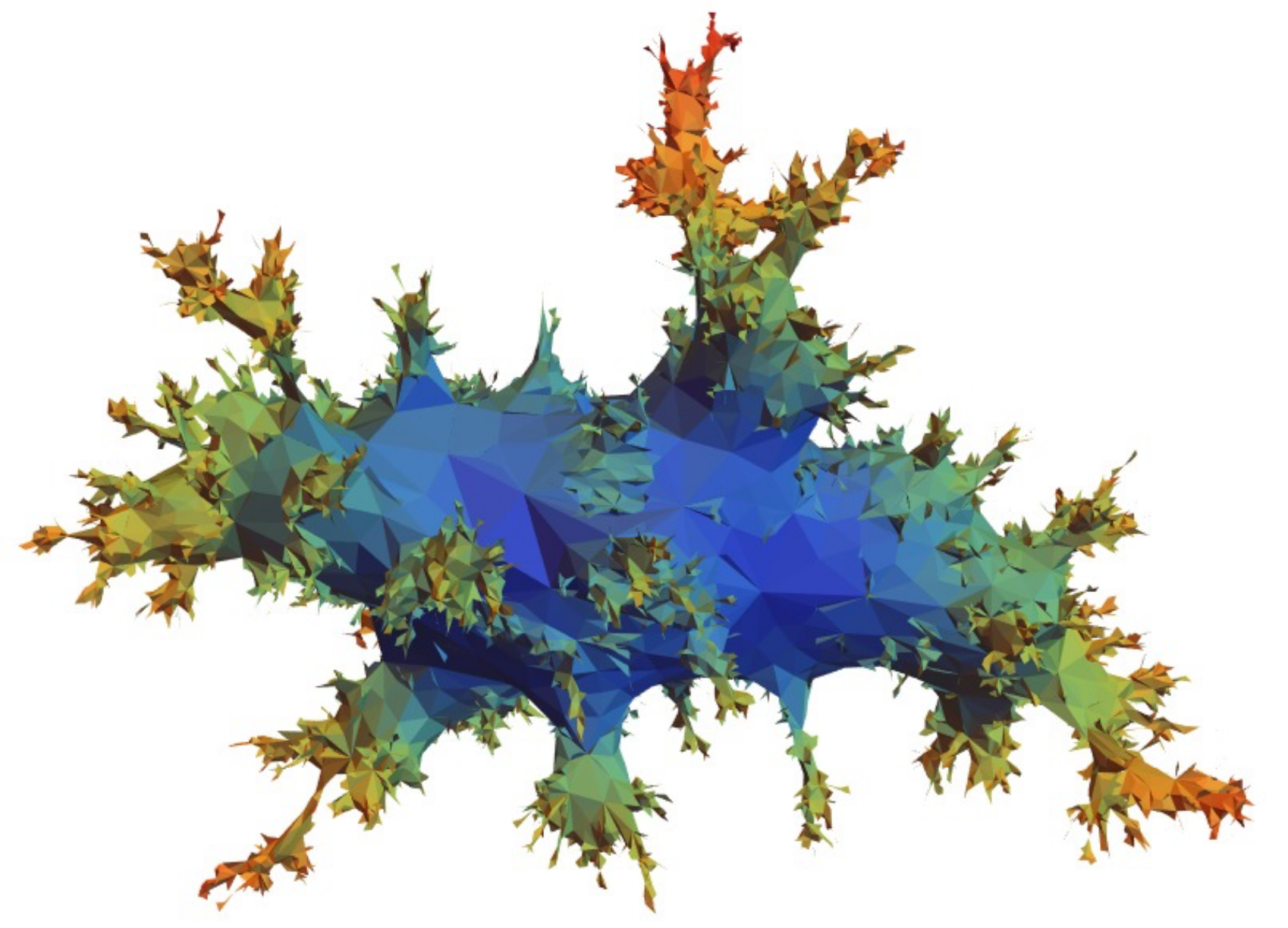}}}}
\caption{Quantum-fluctuating spacetime foam, from a 2D toy model of quantum gravity. (Courtesy of T.\ Budd)}
\label{fig:foam}
\end{figure}

While the continuum limit of a two-dimensional toy version of the path integral (\ref{path}) can be found analytically \parencite{firstcdt}, unsurprisingly for a
strongly interacting quantum field theory in four dimensions this is of course not possible for quantum gravity proper. 
The second key point is that a nonperturbative evaluation of the CDT-regularized path integral and of suitable observables is nevertheless possible 
numerically, by using powerful Monte Carlo methods.
Although the idea of lattice quantum gravity as the gravitational analogue of lattice QCD has been around for a long time, as described in \textcite{livrev}, 
due to the dynamical nature of spacetime geometry and its invariance structure
it has taken much longer to adapt the very successful arsenal of lattice methods in quantum and statistical field theory to the case of gravity, 
culminating in today's lattice gravity \`a la CDT.

The ability to perform numerical experiments allows us to quantitatively explore the dynamical content of the gravitational path integral 
near the Planck scale. In the absence of actual experiments, this provides a unique and valuable reality check on how gravity behaves on ultrashort length
scales, where spacetime is supposed to become a quantum spacetime foam (cf.\ Fig.\ \ref{fig:foam}). It also serves as a concrete blueprint for the type 
of universal results
we can expect quantum gravity to deliver, given the nature of diffeomorphism-invariant observables and the structural lack of detailed analytical control
in a nonperturbative Planckian regime. In what follows, we will highlight key structural properties and results obtained from using Causal Dynamical 
Triangulations; further technical details can be found in the major reviews by \textcite{review1,review2}.

\section{The need for a nonperturbative path integral}
\label{sec:found}

The action that gives rise to the classical Einstein equations,
\beq\label{j2-1}
S[g_{\mu\nu},\phi] = \frac{1}{16 \pi G} 
\int d^4 x \sqrt{|\det g|} \,( R - 2\Lambda)  +  \int d^4 x \sqrt{|\det g|}\ {\cal L }_m ( g_{\mu\nu},\phi),
\eeq
consists of a gravitational part, depending only on the Lorentzian metric field $g_{\mu \nu}$, the 
so-called Einstein-Hilbert action, and a matter part with Lagrangian 
${\cal L}_m$, which includes any matter fields $\phi$ coupled to gravity.
In eq.\ (\ref{j2-1}), $G$ is the gravitational and $\Lambda$ the cosmological coupling constant.
With the standard choice of units $\hbar = c =1$, Newton's constant $G$ has negative mass dimension $-2$.
This implies that a quantum theory of gravity that is based on a perturbative expansion in $G$ around flat spacetime 
will not be a renormalizable quantum field theory. 

To make sense of the gravitational path integral \rf{path}, one needs to define it non\-per\-turbatively.
For nongravitational field theories on flat spacetime, lattice quantum field theories provide 
such a definition. The steps are as follows: Wick-rotate from flat Minkowskian spacetime
to flat Euclidean space by analytically continuing to imaginary time, which will usually result in a well-defined 
Euclidean action. Replace the continuum by a lattice, 
such that its coordinates $x$ are replaced by discrete lattice point $x_i$, and the fields $\phi(x)$ are 
represented by lattice fields $\phi(x_i)$. Then find a suitable lattice action for the fields $\phi(x_i)$ that optimally approximates 
the continuum action. Starting from a lattice with a finite number of lattice points, the lattice-regularized 
path integral will be a finite-dimensional, well-defined integral.
One can then study the limit in which the number of lattice points goes to infinity and the distance between lattice points to zero.
If such a limit can be shown to exist, one has a nonperturbative definition of the 
Euclidean path integral and thus of a corresponding Euclidean quantum field theory. 
To get back to Minkowskian spacetime one can appeal to the Osterwalder-Schrader reconstruction theorem,
which ensures that this continuation is possible under rather general conditions \parencite{mm}.
Lattice field theories have been very useful in addressing nonperturbative 
aspects of renormalizable quantum field theories. When it comes to nonrenormalizable
quantum field theories, they are virtually the only tool we have at our disposal to try to define them in a rigorous way.

In view of this state of affairs, putting gravity on the lattice seems like an obvious strategy to   
make sense of the gravitational path integral \rf{path}, which involves an integration over geometries.
However, one is immediately confronted with a number of problems. 
The program outlined above fails already at the first step:
in the absence of a Minkowskian background geometry and a preferred notion of time, 
there is no obvious analogue of the imaginary-time prescription underlying the standard Wick rotation,   
or known way to associate a general Lorentzian metric with a Riemannian one.  
A popular way to sidestep the lack of a rotation to Euclidean signature is to start from a different 
set of field configurations, namely the space of four-dimensional Riemannian metrics $g^{\rm eu}_{\mu\nu}$.
These are unphysical, in the sense of lacking a causal structure or notion of time. The associated Euclidean path integral 
\begin{equation}
Z^{\rm eu}=\! \int\! {\cal D} [g^{\rm eu}]\; \mathrm{e}^{\, -S^\mathrm{grav}[g^{\rm eu}]}
\label{patheu}
\end{equation}
defines what is usually called \textit{Euclidean quantum gravity}, and has a priori no relation with its Lorentzian counterpart (\ref{path}).  
If we nevertheless use it as input for the lattice theory, we immediately face the problem that the Euclidean 
Einstein-Hilbert action $S^{grav}[g^{\rm eu}]$ is unbounded below, compounding the ill-defined nature of (\ref{patheu}). 
Setting this issue aside still leaves the challenge of finding a lattice discretization which constitutes a good approximation of the space of
all geometries and properly takes the diffeomorphism-invariance of gravity into account. 
Even if we manage to appropriately represent the gravitational lattice fields and action, 
compute the path integral and take the limit of infinite lattice size and vanishing lattice spacing $a$, it is
not obvious how to get back to a quantum field theory of Lorentzian geometries without an Osterwalder-Schrader theorem.

Remarkably, the above steps can be accomplished when one starts from the Lorentzian path integral (\ref{path}),
discretizes the space of geometries rather than the metrics and rotates these to Euclidean signature, before performing the path integral.
This turns out to yield different results from putting Euclidean quantum gravity on the lattice and has taken nonperturbative quantum gravity
to a new level, in developments that will be summarized below.

\section{Simplicial building blocks and setting up the path integral}
\label{sec:quant}

\subsection{Euclidean case}
\label{sec:eubui}

To illustrate the key construction principles of CDT and how they address some of the issues raised in Sec.\ \ref{sec:found},
we first discuss the technically slightly easier case of \textit{Euclidean dynamical triangulations}, abbreviated \textit{DT} or \textit{EDT},
the Euclidean-signature precursor of CDT lattice gravity. On the one hand, it implements the idea of discretizing geometries
rather than metrics, without ever appealing to a coordinate system. On the other hand, it implements the idea of \textit{random geometry}, where
the configuration space of the path integral consists of geometries that can be assembled in all possible ways from a small number of elementary  
building-block types, subject to a set of elementary gluing rules \parencite{book1}.

The building blocks of EDT in $d\! =\! 4$ are four-dimensional equilateral flat simplices, whose flat interior geometry is
determined completely by the geodesic length of their one-dimensional links or edges. Path integral configurations are obtained 
by gluing these four-simplices pairwise along their three-dimensional faces, subject to simplicial-manifold conditions, either with or without
restrictions on the overall topology. A closed equilateral triangulation $T$ is a particular case of a piecewise linear manifold, and a Riemannian
version of classical Regge calculus \parencite{regge} can be used to assign to it a simplicial approximation 
\beq\label{j3-1}
S^{\rm eu}(T) = -\kp_{2} N_{2}(T) + \kp_4 N_{4}(T)
\eeq
of the Einstein-Hilbert action, where $N_k(T)$ denotes the number of $k$-dimensional simplices in $T$.
Assigning a physical, dimensionful length $a$ to the links of the triangulations, the
dimensionless lattice coupling constants $\kp_{2}$ and $\kp_4$ in (\ref{j3-1}) can be expressed in terms of $a$ and the 
previously introduced continuum couplings $G$ and $\Lambda$.
The exceedingly simple form of (\ref{j3-1}) comes from the fact that volumes in equilateral simplicial manifolds come in discrete units, as do
the deficit angles associated with the two-dimensional subsimplices, which are a measure of intrinsic curvature (cf.\ Sec.\ \ref{sec:curv}).  
The Euclidean path integral can then be written as 
\beq\label{j3-2}
Z^{\rm eu}(\kp_{2},\kp_4) = \sum_{T}\tfrac{1}{C(T)}\ \e^{\,\kp_2 N_{2}(T) - \kp_4 N_4(T)} = 
\sum_{N_{2},N_4} {\cal N}(N_{2},N_4) \;  \e^{\,\kp_2N_{2} - \kp_4 N_4} ,
\eeq
where ${\cal N}(N_{2},N_4)$ is the number of triangulations containing $N_4$ four-simplices and $N_2$ triangles, and $C(T)$ is the size of
the automorphism group of $T$.
This illustrates that the partition function (\ref{j3-2}) is purely combinatorial. If we restrict the discrete volume $N_4$ to be finite, $Z^{\rm eu}(\kp_{2},\kp_4)$ 
is well defined
and finite, because there is only a finite number of configurations contributing to the right-hand side. 
If the restriction of finite $N_4$ is lifted and arbitrary topologies for $T$ are allowed, one can show that the growth of 
$ {\cal N}(N_{2},N_4)$ with $N_4$ is such that $Z^{\rm eu}(\kp_{2},\kp_4) $ is ill-defined for any choice of the coupling constants $\kp_{2}$ and $\kp_4$.
One therefore cannot study the limit as $N_4\! \to\! \infty$. The superexponential growth of 
$ {\cal N}(N_{2},N_4)$ can be traced to the proliferation of triangulations with 
a complicated topology. On the basis of numerical evidence, it is believed that
$ {\cal N}(N_{2},N_4)$ only grows exponentially with $N_4$ if the topology of the 
piecewise linear manifolds in the sum \rf{j3-2} is fixed.  

Assuming that this conjecture is correct, let us from now on fix the topology of the contributing geometries. 
Then there exists a so-called \textit{critical coupling} $\kp_4^c(\kp_2)$ such that $Z^{\rm eu}(\kp_{2},\kp_4) $ is well-defined for
 $\kp_4\! >\! \kp_4^c(\kp_2)$ and we can study the large-$N_4$ limit for
 given $\kp_{2}$ by fine-tuning $\kp_4$ to $\kp_4^c(\kp_2)$ from above. 
Using the path integral $Z^{\rm eu}(\kp_{2},\kp_4)$ to calculate 
the average number $\la N_4 \ra$, we want to fine-tune in such a way that
$\la V_4 \ra \propto \la N_4 \ra a^4$ can be interpreted as the average continuum volume of the universe as $a \to 0$.
 
Note that the choice of building blocks has tamed the problem of the unbounded Euclidean action mentioned in Sec.\ \ref{sec:found}. 
The latter can be traced to the conformal mode of the metric, whose kinetic term enters the action with a negative sign. It means that 
ever more rapid oscillations of this mode can make the action arbitrarily large and negative. 
Using identical building blocks imposes a uniform limit on how rapidly the geometry can 
oscillate, set by the size of the simplicial building block, which implies that the regularized path integral has no conformal divergence.
An analysis of the phase diagram of the system shows that the suppression of the unboundedness of the action persists also in the
continuum limit $N_4\!\to\!\infty$, $a\!\to\! 0$, as long as the bare coupling constant $\kappa_4$ is chosen large enough, 
$\kp_4 > \kp^c_4(\kp_{2})$. 

To summarize, the regularized Euclidean path integral \rf{j3-2} can be viewed as the partition function of 
a statistical system of piecewise linear geometries constructed from elementary, triangular
building blocks, where the name \textit{dynamical triangulations} emphasizes the fact that unlike in standard lattice field theory,
the lattice is not fixed but itself plays the role of a statistical field. 
One can then proceed to use methods from the theory of critical phenomena to study the system's infinite-volume limit.
This can be done analytically for the analogue in $d\! =\! 2$ of the path integral \rf{j3-2}, reproducing the results of an evaluation of
the Euclidean path integral in the continuum (see \textcite{book1} for a review).

However, it has not been possible to find interesting continuum limits of the EDT path integral \rf{j3-2} in $d\! =\! 4$.
For small $\kp_2$, typical triangulations are completely crumpled, in the sense that their four-simplices are clustered 
around a few lattice vertices of extremely large order, 
while triangulations at large $\kp_{2}$ degenerate into so-called \textit{branched polymers} or \textit{trees}. 
A phase transition at $\kp_{2} = \kp_{2}^c$ separates these two phase space regions of degenerate 
geometry, but unfortunately it is a first-order phase transition. As a result, 
there is no smooth transition between the two phases and no obvious candidate for taking a continuum limit. Efforts to locate 
a second-order transition in Euclidean quantum gravity by using a more complicated action than \rf{j3-1} are ongoing \parencite{higher-d-edt}.

 \subsection{Lorentzian case}
 \label{sec:lorbui}
 
\begin{figure}[t]
\centerline{\scalebox{0.5}{\rotatebox{0}{\includegraphics{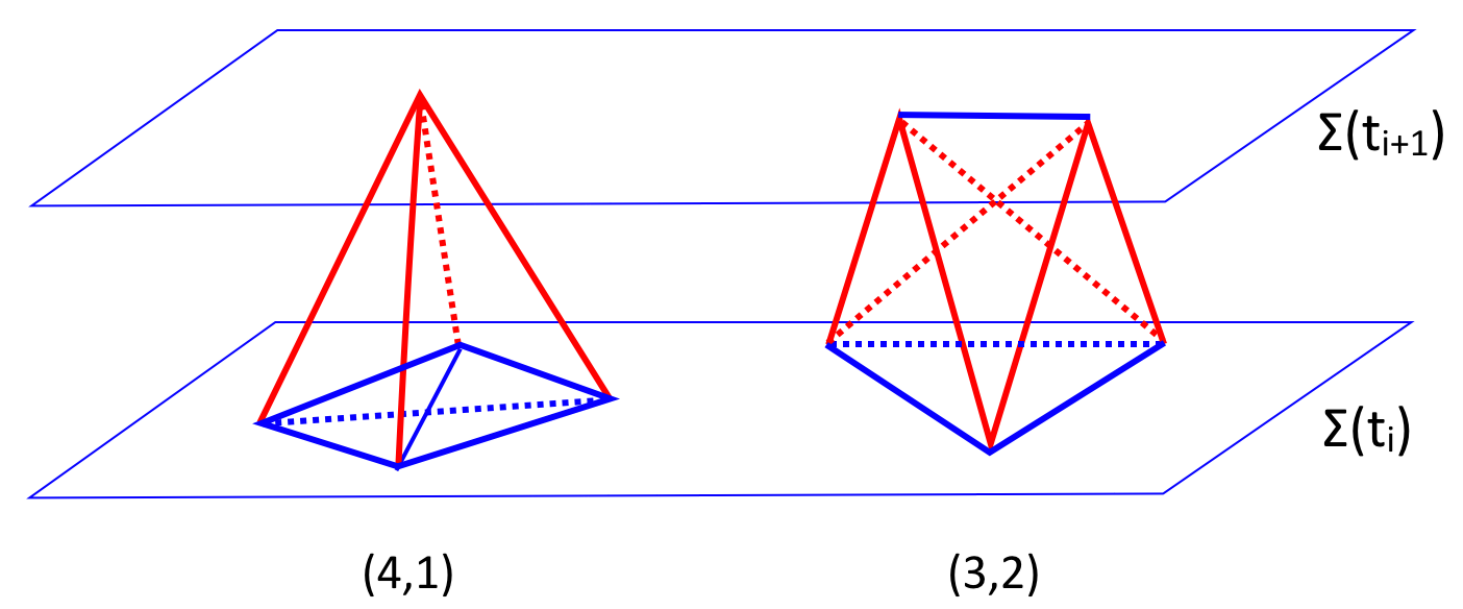}}}}
\caption{CDT building blocks of type (4,1) and (3,2) interpolating between spatial
slices $\Sigma(t_i)$ and $\Sigma(t_{i+1})$; spacelike edges in blue and timelike ones in red.  }
\label{figj1}
\end{figure}

Quite apart from these unresolved issues of the nonperturbative Euclidean path integral, our primary interest
is of course physical, Lorentzian quantum gravity, obtained from the original path integral \rf{path} over Lorentzian geometries.
The method of causal dynamical triangulations employs a Lorentzian analogue of the equilateral building blocks of EDT,
while ensuring the presence of a well-defined analytic continuation of the complex path integral to a real partition function that
can be investigated with the help of Monte Carlo simulations. A choice of four-simplices which realizes these objectives is
shown in Fig.\ \ref{figj1}. The interior geometry of the simplices is that of flat Minkowski space, and their edges come in two different types, 
namely space- and timelike, with two different squared edge lengths
\beq\label{j3-3}
\ell^2_s = a^2, \qquad \ell^2_t = -\alpha a^2, \quad \alpha > 0,
\eeq
where $a$ again denotes the lattice spacing (and short-distance regulator) and $\alpha$ is a constant. Unlike in the Riemannian situation of
Sec.\ \ref{sec:eubui}, each four-simplex now has a well-defined causal (or lightcone) structure. 

The other ingredient that must be specified to implement the regularized path integral are the gluing rules for these building blocks. 
The gluing should again yield a simplicial manifold.
In addition, only faces of the same type, with respect to the space- and timelike character of their edges, can be glued together. 
The most important difference with respect to the Euclidean case is that each triangulation must
have a well-behaved causal structure, in the sense of obeying a lattice version of global hyperbolicity. This means not only a fixed topology of spacetime,
but a topology that is the direct product $M \! =\! [0,1]\times \Sigma$ of a time direction and a fixed spatial topology $\Sigma$. 
This is implemented by giving each four-dimensional triangulation $T$ a sliced structure, such that $T$ is given by a sequence of three-dimensional 
spatial triangulations $\Sigma({t_i})$, labelled sequentially by an integer $t_i$ representing a discrete notion of proper time. A more detailed
discussion of the nature and physical interpretation of this notion of time is given in \textcite{review2}. 

Note furthermore that the presence of
a notion of time is essential to proving unitarity of the continuum theory, 
by showing that the transfer matrix with respect to this time obeys reflection positivity. This is
nontrivial to achieve in lattice gravity and systems of random geometry in particular, as reviewed by \textcite{livrev}, but has been
shown in CDT (see \textcite{dyn,review1} for definitions and further discussion).

Note that a priori there are no hard physical arguments for what kind of configurations should be included in the nonperturbative gravitational 
path integral, and folklore from nongravitational theories may not be applicable. The choice to impose strong conditions on the causal 
structure of individual path integral histories is in the spirit of what has been suggested by \textcite{teitelboim} in a continuum context. 
The relevant hurdles that must be cleared by any alternative prescription is to show that the regularized path integral exists and is well-defined, and that an
appropriate continuum limit can be taken, leading to interesting physical results and reproducing classical features in an appropriate
classical limit. Lattice gravity based on CDT passes these nontrivial tests and at the same time is the only known cure to terminal pathologies
like the geometric crumpling and polymerization seen in EDT.  

Each slice of constant integer time is a three-dimensional equilateral Euclidean triangulation with edges of length $a$, in accordance with (\ref{j3-3}). 
The spacetime between two subsequent slices $\Sigma(t_i)$ and $\Sigma(t_{i+1})$ consists of a layer of four-simplices of type
$(m,5\mi m)$, $m=1,\ldots,4$, which are defined as sharing a $(m\mi 1)$-simplex with $\Sigma(t_i)$
and a $(4\mi m)$-simplex with $\Sigma(t_{i+1})$. Fig.\ \ref{figj1} shows the four-simplices of type (4,1) and (3,2); those of type (1,4) 
and (2,3) can be obtained by time reflection. As indicated, the links connecting neighbouring spatial slices are timelike, while all others are
spacelike.

The simplicial form $S(T)$ of the Einstein-Hilbert action obtained from Regge calculus is again simple and linear in the counting variables
$N_4^{(m,5-m)}(T)$ of the various four-simplex types, as well as the number $N_0(T)$ of vertices. For a closed $T$, whose time direction has
been compactified for simplicity, the action is
\beq\label{j3-4a}
S(T) = \tilde{k}_0 N_0(T)+\tilde{k}_1 \Big( N_4^{(1,4)}(T)+N_4^{(4,1)}(T)\Big) +\tilde{k}_2 \Big( N_4^{(2,3)}(T)+N_4^{(3,2)}(T) \Big),
\eeq
where the coupling constants $\tilde{k}_i$ depend linearly on the lattice couplings $\kp_{2}, \kp_4$ introduced earlier, and nonlinearly on the parameter $\alpha$
(see \textcite{dyn, review1} for details). The regularized version of the Lorentzian path integral \rf{path} now reads
\beq\label{j3-4b}
Z(\kp_{2},\kp_4;\alpha) = \sum_{T}\tfrac{1}{C(T)}\  \e^{\, i S(T)},
\eeq
where the summation is over causal triangulations assembled according to the gluing rules discussed above. 
For a finite upper bound on the discrete volume $N_4$, (\ref{j3-4b}) is finite and well defined. 
 
An added bonus of this construction is that for each Lorentzian triangulation with length assignments (\ref{j3-3}) 
we can perform an analytic continuation in the lower half of the complex $\alpha$-plane 
such that $\alpha$ changes from positive to negative values. This turns all timelike edges into spacelike ones, and the corresponding 
piecewise flat geometry becomes Euclidean. For a certain range of $\alpha$, including 
$\alpha\! =\! 1$, we obtain the usual relation between the Euclidean and Lorentzian actions, namely
\beq\label{j3-5}
S(T,-\alpha \mi i\epsilon) =: iS_E(T,\alpha), \qquad \ell^2_t = \alpha a^2 > 0,
\eeq   
where the symbol $S_E$ denotes the Euclidean action after the analytic continuation.
It implies that we can compute the path integral using the real Boltzmann weights 
$\e^{\, -S_E(T)}$ instead of the complex weights $\e^{\.i S(T)}$, and after performing 
the sum make the analytic continuation $-\alpha \to \alpha$ back to the  
Lorentzian path integral. Note that the set of piecewise flat 
Euclidean geometries used in the construction of this analytically continued path integral 
is a strict subset of the Euclidean geometries considered in EDT, which will be seen to 
lead to different and much more interesting results.
To summarize, 
\beq\label{j3-6}
Z(\kp_{2},\kp_4;\alpha) \to Z(\kp_{2},\kp_4;-\alpha)=
Z_E(\kp_{2},\kp_4;\alpha) = 
\sum_{T}\tfrac{1}{C(T)} \ \e^{\, - S_E(T, \alpha)},
\eeq
where 
\beq\label{j3-7}
S_E(T,\alpha) = \hat{k}_0 N_0(T)+\hat{k}_1 \Big( N_4^{(1,4)}(T)+N_4^{(4,1)}(T)\Big) +\hat{k}_2 \Big( N_4^{(2,3)}(T)+N_4^{(3,2)}(T) \Big),
\eeq
with $\hat{k}_a(\alpha)\! :=\! -i \tilde{k}_a (-\alpha)$.
Like in the EDT case discussed in Sec.\ \ref{sec:eubui}, the gravitational path integral $Z_E$ has the form of a statistical partition function of
random geometries.
A two-dimensional version has been solved analytically by \textcite{firstcdt}, leading to results that are inequivalent to those of the Euclidean 
two-dimensional path integral, with a different universal behaviour of its observables, and a natural analytical continuation of
the obtained continuum propagator back to Lorentzian signature. Although these are mere toy models of the four-dimensional
theories, they highlight the inequivalence between Lorentzian and Riemannian metric signature in nonperturbative quantum gravity.  

The analysis of the four-dimensional Lorentzian path integral in terms of a CDT lattice formulation and its results to date are the subject of
the following sections. A crucial tool for extracting quantitative results beyond perturbation theory are Markov chain Monte Carlo methods. 
Their application to systems of dynamical geometry presents specific challenges, whose resolution has been described elsewhere, see e.g.\ 
\textcite{review1}.

\subsection{Including matter fields}

We have so far considered the lattice formulation of the gravitational, Einstein-Hilbert
part of the action \rf{j2-1}, but the inclusion of matter in this formalism is straightforward. 
To illustrate the point, let us discuss how to incorporate a scalar field $\phi(x)$ with continuum (Euclidean) action
\beq\label{j3-10}
S_m[\phi,g_{\mu\nu}] = \int_{M} d^4 x \sqrt{\det g} \;\Big( g^{\mu\nu} 
\partial_\mu \phi(x) \partial_\nu \phi(x) + V(\phi) \Big). 
\eeq
The simplest CDT lattice implementation of this action on a triangulation $T$ is obtained by placing the scalar fields 
at the vertices $x_i$ of the lattice dual to $T$ (equivalently, at the centres of the four-simplices of $T$)
and using as lattice derivative the difference $\phi(x_i) -\phi(x_j)$ between 
scalar fields at neighbouring pairs of such vertices. For a given $T$ this leads to the lattice action
\beq\label{j3-11}
S_m[\phi,T] = \sum_{ (x_i,x_j)} \big(\phi(x_i) -\phi(x_j)\big)^2 +
\sum_{x_i} V(\phi (x_i)),
\eeq
where $(x_i,x_j)$ denotes pairs of neighbouring vertices, i.e.\ links in the dual lattice. 
To obtain a combined gravity-matter path integral, one integrates for each triangulation $T$ over all matter field configurations
$\{ \phi(x_i)\}$, weighted by $\e^{-S_m[\phi,T]}$, and then perform the sum over all geometries.
On both EDT and CDT lattices one can also incorporate gauge fields \parencite{gaugefields1,gaugefields2},
as well as fermions. The latter can be included because local frames can be introduced on individual four-simplices in a straightforward
way as shown explicitly by \textcite{wilsonloop} in their analysis of gravitational Wilson loops.

\section{CDT as a lattice and continuum field theory}
\label{sec:lattcont}

Having specified the ingredients which turn the formal Lorentzian path integral (\ref{path})
into a properly regularized expression, based on a statistical system of piecewise linear
geometries with a well-defined causal structure, we can tap into 
standard concepts and techniques from the theory of critical phenomena 
to understand how this lattice system can give rise to a continuum field theory.

The first thing to stress is that the chosen building blocks do not have a fundamental status, and do 
not define a fundamental physical length scale. On the contrary, the link length $a$
plays the role of a quantum field-theoretic UV cutoff. To define an interacting continuum
theory of quantum gravity it must be taken to zero.
Taking a Wilsonian point of view on the renormalization of lattice field 
theories, we should look for second-order phase transitions of the lattice-regularized model. 
Such transitions are associated with divergent correlation 
lengths of the lattice fields, which implies that the physics related to these correlation 
lengths will exhibit \textit{universality}: it will to a large degree be independent of the arbitrary
discretization choices made when setting up the lattice theory, and many different lattice theories can lead 
to the same continuum quantum field theory, if it can be shown to exist.

Explicit universality has been observed in two-dimensional toy versions of the path integral, and is expected
to be present in four dimensions also. The two-dimensional models can be solved 
analytically in either signature, 
as has already been mentioned, and provide beautiful illustrations of Wilson's ideas. For example, one can
vastly extend the space of configurations and coupling constants by including not only triangles, but also
squares, pentagons, etc. with different weights. The enlarged phase space contains 
a critical surface of finite co-dimension where one can take 
the cutoff $a\! \to\! 0$ such that a suitable correlation length diverges, leading to  
what one can view as continuum theories of
two-dimensional quantum gravity, see \textcite{book1,book2} for details. 
The important message from these exactly soluble examples is that the lattice formalism of dynamical triangulations
provides a natural regularization of diffeomorphism-invariant, backgound-independent 
quantum theories of gravity and geometry, without any need to question the foundational principles of
quantum field theory.

\subsection{Phase diagram and phase transitions}
\label{sec:phasediagram}

Turning next to the evaluation and analysis of the gravitational path integral formulated in terms of CDT, 
we follow the strategy of looking for second-order phase transitions in the phase space spanned by the
bare coupling constants. The Wick-rotated action $S_E$ in \rf{j3-6} and \rf{j3-7} depends on the three parameters $\kp_2,\kp_4$ and 
$\alpha$. One can debate to what extent $\alpha$ can be viewed as a coupling 
constant, since it was introduced merely as a finite scale-factor between space- and timelike link lengths.
However, the results reported below suggest that in a Wilsonian spirit it is 
more appropriate to view the resulting $\alpha$-dependence as a generalization of the 
Regge action for $\alpha\! =\! 1$, which classically is an equally valid discretization of the Einstein-Hilbert action. 
The conventional parametrization of the action \rf{j3-7} is given by
\beq\label{j4-1}
 S_E(T; k_0,k_4,\Delta) =-\big( k_0\,\plu\, 6 \Delta\big) N_0(T) +
 k_4 N_4(T)+ \Delta \big( N_4^{(4,1)}(T) \plu N_4^{(1,4)}(T)\big),
 \eeq
 where
 \beq\label{j4-2}
 N_4(T) := N_4^{(1,4)}(T)\plu N_4^{(4,1)}(T)\plu  N_4^{(3,2)}(T)\plu  N_4^{(2,3)}(T)
 \eeq
and the coupling constants $k_0$,  $k_4$ and $\Delta$ are specific functions of $\kp_2,\kp_4$ and $\alpha$,
where we again refer to \textcite{review1} for details.
In particular, the \textit{asymmetry parameter} $\Delta$ satisfies $\Delta\! =\! 0$ for $\alpha\! =\! 1$.
 
Since the corresponding four-dimensional path integral or partition function $Z_E(k_0,k_4,\Delta)$ cannot be calculated analytically, 
the phase diagram of the theory is explored with the help of Monte Carlo simulations. 
For the results reported below the topology chosen for the spatial slices was spherical, $\Sigma\! \cong\! S^3$.
There are also simulations for toroidal slices of topology $T^3$, which have been reviewed by \textcite{torusreview}.

\begin{figure}[t]
\centerline{\scalebox{0.8}{\rotatebox{0}{\includegraphics{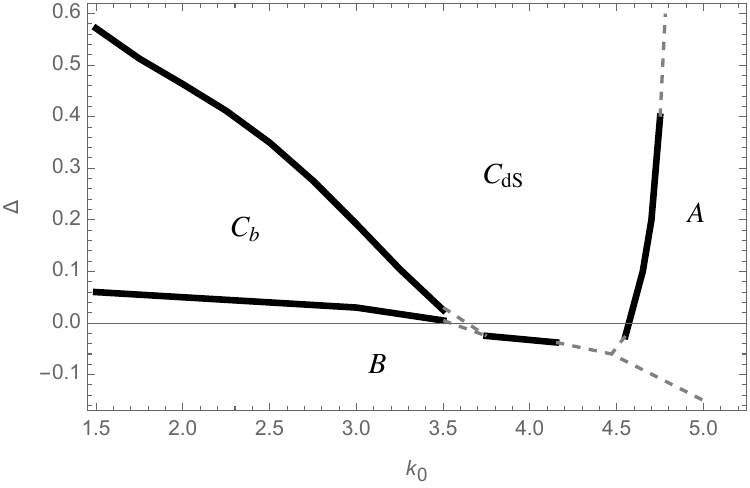}}}}
\caption{Phase diagram of CDT lattice gravity.}
\label{figj2}
\end{figure}

The phase structure for the spherical case has the following features.  
Analogous to what was reported in Sec.\ \ref{sec:eubui} for the coupling $\kappa_4$ in EDT, 
there is a transition from a finite to an infinite expectation value $\la N_4 \ra$ of the four-volume at 
a critical value $k_4^c(k_0,\Delta)$ of $k_4$, and
the partition function is not defined for $k_4 < k_4^c(k_0,\Delta)$.
As a consequence, there is a two-dimensional critical surface parametrized by $k_0$ and $\Delta$. 
Since we are interested in the behaviour for large $\la N_4 \ra$, the simulations should stay as close to this surface as possible.
Note that in itself, the \textit{infinite-volume} or \textit{thermodynamic limit} $\la N_4 \ra \to \infty$ does not  
necessarily imply interesting continuum physics. 
For that, we need to scan the two-dimensional coupling-constant plane spanned by $(k_0,\Delta)$ for the
presence of phase transitions of second order.
The result of this analysis is shown in Fig.\ \ref{figj2}.
 
Of the four phases found, the \textit{de Sitter phase} $C_{\rm dS}$ has our main interest, as will become clear
below. Two of the other phases can be thought of as Lorentzian analogues of the degenerate phases of
EDT, namely of the branched-polymer phase dominated by the conformal mode (phase $A$), and
of the crumpled phase (phase $B$). The bifurcation phase $C_{\rm b}$ is characterized by the emergence of
a distinguished, string-like structure consisting of vertices of high order, indicating a breakdown of the
conjectured homogeneity of geometry in the de Sitter phase \parencite{signature}. While the physical relevance of the 
$C_{\rm b}$-phase is still being explored, there is strong evidence that interesting physical behaviour 
is present in the $C_{\rm dS}$-phase, regarding both infrared and ultraviolet properties. This also means
that the phase transition lines bordering the de Sitter phase are of particular interest. 
The $C_{\rm dS}$-$C_{\rm b}$ transition line has been identified as second-order \parencite{exploring},
and the $C_{\rm dS}$-$B$ transition line is another promising candidate under investigation.
Points along these transition lines are prime candidates in the search for UV fixed points at which one may be able to find 
a continuum theory that is well-behaved in the UV.   
Before reporting on progress in locating such a fixed point, we will first describe some of the infrared physical properties 
that have been found in the de Sitter phase.

\subsection{The quantum de Sitter universe}
\label{sec:QdSU}

One of the remarkable features of the gravitational path integral in terms of CDT is the emergence
of a macroscopically four-dimensional quantum spacetime whose large-scale properties are compatible with those of a
classical de Sitter universe. We start by reviewing the evidence that the dynamically generated quantum spacetime is de Sitter-shaped,
as shown by \textcite{desitter1,desitter2}.

The numerical simulations take place inside the de Sitter phase and
for com\-pu\-ter-technical reasons are performed such that
the discrete four-volume $N_4$ of the geometries fluctuates around a given value $\bar{N}_4$. 
Running Monte Carlo simulations at different and increasing volumes $\bar{N}_4$ then allows for the 
use of powerful finite-size scaling methods when analyzing the data.
In addition, the computational set-up involves a fixed number $t_{\rm tot}$ of time steps, but it turns out that
the precise choice is unimportant as long as $t_{\rm tot}$ is sufficiently large. 

\begin{figure}[t]
\centerline{
{\scalebox{0.7}{\includegraphics{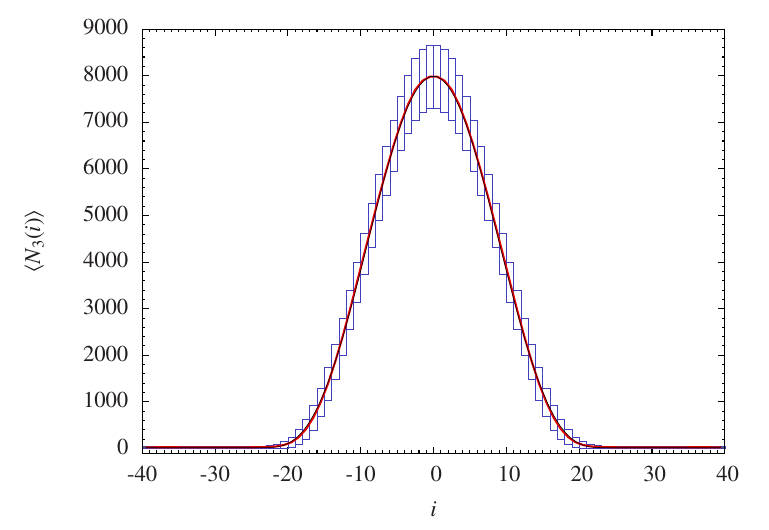}}}
}
\caption{Volume profile and fluctuations as a function of time $i$. Within the range of nonvanishing three-volume, the measured volume profile 
$\langle N_3(i)\rangle_{\bN_4}$ at $\bN_4 = 360.000$, extrapolated to continuous time (red curve) and the $\cos^3$-profile on the right-hand side 
of eq.\ \rf{j5-1} (black curve) are indistinguishable at the given resolution. Blue bars indicate the size of fluctuations $\Delta N_3 (i)$ around the average.  
}
\label{figj3}
\end{figure}

In the simulations one can easily monitor the number $N_3(i)$ of three-simplices in a 
three-dimensional spatial triangulation at a given time $t_i$. The function $N_3(i)$ is called the shape or
\textit{volume profile} of the spacetime. We can measure the
expectation value $\la N_3(i) \ra_{\bar{N}_4}$ and
the correlator $\la N_3(i) N_3(j) \ra_{\bar{N}_4}$, as well as the corresponding fluctuations of $N_3(i)$. 
The outcome is illustrated in Fig.\ \ref{figj3}.
The volume profile consists of an extended part of nonvanishing macroscopic volume and a stalk,
where the three-volume is as small as allowed by the regularity condition imposed (i.e.\ $N_3(i)\! =\! 5$) to prevent the universe from pinching off. 
In other words, spatial volume in the stalk is zero from a continuum point of view. 

The beautiful result is that the overall shape observed for the nonvanishing part of the volume profile is described perfectly by the 
functional expression
\beq\label{j5-1}
\la N_3(i) \ra_{\bN_4} = c \,\bN_4 \; \frac{3}{4}\, \frac{1}{\omega \bN_4^{1/4}} \, 
\cos^3 \bigg( \frac{i}{\omega \bN_4^{1/4}} \bigg),
\eeq 
and the correlator of the three-volume fluctuations by
\beq\label{j5-2}
\la \delta N_3(i) \delta N_3(j) \ra_{\bN_4} = 
\bN_4\; F\bigg( \frac{i}{\omega \bN_4^{1/4}},
\frac{j}{\omega \bN_4^{1/4}} \bigg) ,\quad \delta N_3(i) := N_3(i)\mi \la N_3(i) \ra ,
\eeq
where $c$ and $F(0,0)$ are of order 1, and $c$ and $\omega$ are 
functions of $k_0$ and $\Delta$.
Eq.\ \rf{j5-1} should be compared to the formula for the volume $V_3(t)$
of the three-sphere located at a geodesic distance $t$ from the north pole of a round four-sphere, which is given by
\beq\label{j5-3}
V_3(t) = V_4\; \frac{3}{4} \;\frac{1}{\omega_0 V_4^{1/4}}\;
 \cos^3 \bigg( \frac{t}{\omega_0 V_4^{1/4}}\bigg), \quad \omega_0^4 = \frac{3}{8 \pi^3}.
 \eeq
One concludes that by a suitable finite rescaling of the lattice distance units of space and time, the volume profile generated in
the lattice simulations matches precisely that of the (Euclideanized) classical de Sitter solution to general relativity, anywhere 
in phase $C_{\rm dS}$. Defining $(\Delta N_3(i)_{\bN_4})^2\! :=\! \la \delta N_3(i) \delta N_3(i) \ra_{\bN_4}$ and
assuming one can write $V_3(t) \propto N_3(i) a^3$, we have
\beq\label{j5-4}
\frac{\Delta V_3(t)}{V_3(t) }\Big|_{V_4} = \frac{\Delta N_3(i)_{\bN_4}}{\la N_3(i)\ra_{\bN_4}} .
\eeq
The right-hand side of this relation can be obtained from the Monte Carlo results for \rf{j5-1} and \rf{j5-2}, and has
been found to agree well with the left-hand side, which can be computed using semiclassical perturbation theory around the classical de Sitter solution.
Note that the height of the curve in Fig.\ \ref{figj3} scales like $\bN_4^{3/4}$ and its width like
$\bN_4^{1/4}$, whereas the fluctuations $\Delta N_3 (i)$ according to \rf{j5-2} only scale like $\bN_4^{1/2}$. It means that 
the relative size of the fluctuations scales to zero with increasing four-volume. 

We conclude that CDT lattice gravity, without putting in any 
background geometry, dynamically generates a quantum spacetime whose shape is that of a
semiclassical de Sitter space, including quantum fluctuations, which is an unprecedented result in nonperturbative quantum gravity.

\subsection{The CDT effective action}
\label{sec:effect}

From the knowledge of the measurements of the volume profile $\la N_3(i) \ra$ and the correlator $\la \delta N_3(i) \delta N_3(j) \ra$
one can determine an effective action for $N_3(i)$ that reproduces these measurements. This reconstruction can
proceed in a number of ways, but the final result is amazingly simple, see \textcite{review1,transfermatrix} for details. 
Considering only the physics of the region of nonvanishing three-volume, one obtains
\beq\label{j6-1}
S_{\rm eff}(\bN_4) = \frac{1}{\Gamma}\ \sum_i  \left(
\frac{\big(N_3(i\plu 1) -N_3(i)\big)^2}{ N_3({i})\big)} + \delta \, N_3^{1/3}(i) \right).
\eeq
The quantities $\Gamma$ and $\delta$ depend on $k_0$, $\Delta$ and $\bN_4$, and their dependence on $\bN_4$
becomes stronger as the second-order $C_{\rm dS}$-$C_{\rm b}$ phase transition is approached. For fixed $(k_0,\Delta)$, the
$\bN_4$-dependence disappears for sufficiently large $\bN_4$. 
Introducing $s_i\! =\! i/\bN_4$ and $n_3(s_i)\! =\! N_3(i)/\bN^{3/4}_4$ and replacing the summation over $i$ 
by an integral over $s$ for large $\bN_4$, one can rewrite \rf{j6-1} as
\beq\label{j6-2}
S_{\rm eff} (\bN_4) = 
\frac{\sqrt{\bN_4}}{\Gamma} \int ds \; 
\left(\frac{\dot{n}_3^2(s)}{ n_3(s)} + \delta \, n_3^{1/3}(s) \right).
\eeq   
This is precisely the minisuperspace action of \textcite{hartle-hawking} if one makes the identifications
\beq\label{j6-3}
\frac{\sqrt{\bN_4}}{\Gamma} = \frac{\sqrt{V_4}}{24 \pi G},\qquad 
\delta = \delta_0 \equiv \frac{3}{8 \pi^2}.
\eeq
Although the measured $\delta$ is not equal to $\delta_0$, a rescaling of the 
relative length assignments of space- and timelike links allows us to obtain
such an agreement for any values of $k_0$ and $\Delta$ in the de Sitter phase $C_{\rm dS}$,
as stated earlier.
 
Hartle and Hawking resorted to a minisuperspace reduction of gravity 
to study some aspects of quantum gravity, and the unboundedness problem of the Euclidean action in particular. 
As mentioned in Sec.\ \ref{sec:lorbui} above, CDT provides a nonperturbative solution to this problem. Moreover,
by integrating out all degrees of freedom except for the three-volume $V_3(t)$ in CDT, 
one is effectively deriving the Hartle-Hawking minisuperspace action from full quantum gravity. It would 
be very interesting to determine corrections to the effective action \rf{j6-1}, but the accuracy of the numerical data 
does not currently allow this. 

The above analysis suggests the presence of a semiclassical de Sitter-like universe, unless we are too close to the phase boundaries. 
We can therefore use relation \rf{j6-3} to obtain an estimate of the lattice spacing $a$ in terms of physical length units,
assuming $V_4 \propto \bN_4\, a^4$.
This yields $a\! =\!  \tilde{c} \sqrt{G}$, where $\tilde{c}$ is of order 1. 
In other words, $a$ is of the order of the Planck length, but 
not much smaller, and the typical linear size of a quantum spacetime simulated in the 
computer is in the range of 10---20 Planck length.

Returning to our earlier discussion about the phase structure, we are interested in identifying and approaching a possible UV fixed point. 
Using various renormalization group techniques, such a fixed point has been found in continuum quantum gravity
\parencite{renormalizationgroup}. We have seen that in CDT lattice gravity any putative fixed point 
will likely be located on one of the boundaries of the de Sitter phase $C_{\rm dS}$. The effective 
action \rf{j6-2} can be useful in relating the two formulations, since it can be compared to the effective actions 
calculated in the renormalization group analysis. The quantity $\Gamma$ measured in CDT changes 
with $k_0$ and $\Delta$, and close to the boundary also with 
$\bN_4$, while the quantities $G$ and $V_4$ on the right-hand side of eq.\ \rf{j6-3}, calculated  
with the help of the renormalization group, changes as a function of the renormalization group scale. 
Close to the UV fixed point the latter can be related to the UV cutoff of the lattice theory, given by the lattice spacing $a$.
Work currently in progress indicates that a putative UV lattice fixed point
may be located at one of the two triple points of the CDT phase diagram.

\section{Observables}
\label{sec:obs}

The physical content of quantum gravity is encoded in its geometric observables, which in the context of CDT lattice gravity are given by 
operationally well-defined, diffeomorphism-invariant quantum operators, whose eigenvalues are computable and finite in a continuum limit.
Observables in pure gravity are nonlocal, and in the absence of a reference system -- which could take the form of coupled matter fields or (asymptotic)
boundaries with a fixed metric structure -- are typically given as spacetime integrals or averages of quasi-local
scalar quantities. 

What should be kept in mind regarding the type of observable accessible in a nonperturbative and strongly
quantum-fluctuating regime is the fact that in a continuum limit, typical geometric configurations in the path integral are continuous,
but nowhere differentiable. In particular, they are not smooth, and even finite piecewise flat triangulations do not
possess local coordinate systems that extend over more than pairs of adjacent simplices, due to their singular curvature
structure. There is therefore no useful notion of tensor calculus, and consequently no
meaningful way to talk about a quantum metric $\hat g_{\mu\nu}$ or its expectation value -- disregarding for the moment its noninvariant behaviour under diffeomorphisms --
not even in a coarse-grained or approximate sense. Even when considering local scalar quantities, say, by trying to implement local invariants
of the Riemann tensor in terms of na\"ive finite-difference expressions on the lattice, they will typically be highly divergent in a continuum limit. 
Indeed, a well-defined renormalized notion of Ricci curvature, constructed from quasi-local distance and volume data, has only recently been implemented, 
see Sec.\ \ref{sec:curv} below.
 
These properties are not flaws or peculiarities of this particular formulation of nonperturbative quantum gravity, but reflect the highly nonclassical nature of 
quantum spacetime in a near-Planckian regime. Nevertheless, a well-defined metric
structure is still present, which allows us to measure distances and volumes, and is 
crucial in the construction of quantum observables. In fact, these \textit{rods and clocks of quantum geometry} are close in spirit to how we explore
the properties of classical gravitating systems in our universe with the help astrophysical measurements. 

Secs.\ \ref{sec:QdSU} and \ref{sec:effect} above illustrated the global nature of the observable studied, in this case the total spatial volume $V_3(t)$ as 
a function of proper time $t$, with the latter defined invariantly as the distance from the beginning of the universe at vanishing spatial volume. The robust
character of this observable clearly contributes to the fact that we can match its expectation value with great accuracy to a classical de Sitter volume
profile, and that quantum fluctuations are relatively small in size. For a general quantum observable, there is no a priori guarantee 
that a semiclassical limit will be observable in the scale window that can be accessed computationally. Even more interesting from a quantum-gravitational 
point of view, we may observe genuine quantum signatures, characteristic deviations from an expected classical behaviour that need not be small,
and may not be visible in any perturbative approach. An important example is the spectral dimension of spacetime. It is one of the selected quantum
observables we discuss below. Further examples and more comprehensive treatments of observables can be found in \textcite{review1,review2,curvreview}.

\subsection{Dimensions}
\label{sec:spectral}

In classical gravity, the dimension of spacetime is given by its topological dimension, which is nondynamical and fixed to 4. 
There are various ways of characterizing the dimension of a nonsmooth, nonclassical metric space, which on such a space may not coincide with the
topological dimension or with each other. A powerful and popular method to quantify the behaviour of such a geometry 
is by its spectral and Hausdorff dimension, defined in the form of scaling exponents that characterize the growth of volumes 
(of a diffusion cloud, of a geodesic ball) as function of a diffusion time and a radial distance respectively. 

The relevance of these \textit{fractal dimensions} as observables in quantum gravity is two-fold, as necessary conditions
for the existence of a classical limit and by exhibiting true quantum features. On a smooth manifold, the fractal dimensions are equal to the topological dimension, 
but this is generally not true for quantum geometries, like those obtained from a nonperturbative gravitational path integral.
In particular, even when the microscopic building blocks used to assemble the path integral histories are four-dimensional, the resulting 
quantum superposition will in general not be four-dimensional, on any scale. This somewhat surprising fact was one of the lessons of
Euclidean quantum gravity \`a la EDT, and took a while to recognize. Noncanonical values of fractal dimensions on Planckian scales can be
perfectly acceptable as \textit{quantum signatures}, that is, as expressions of the nonclassical character of the geometry. 
However, the absence of four-dimensionality  
on large scales signals the absence of a well-defined classical limit, making the quantum gravity candidate theory ineligible. 

Spacetime-averaged fractal dimensions can be defined and evaluated on the ensemble of
random geometries underlying the CDT approach, with interesting results. 
Starting with a continuum analysis of the spectral dimension, let $ \Delta_g $ denote the Laplacian on  
a smooth $d$-dimensional Riemannian manifold $M$ with metric $g_{\mu \nu}(x)$. 
Diffusion on $M$ is described by the equation
\beq\label{j5-5}
\frac{\partial}{\partial \sigma} K_g(x,y;\sigma) = \Delta_g(x) K_g(x,y;\sigma),
\eeq
whose solution $K_g(x,y;\sigma)$, the so-called heat kernel,  
is interpreted as the probability that a particle starting at the point $x$ diffuses
(performs a random walk) to the point $y$ in diffusion time $\sigma$. For equal arguments, 
$K_g(x,x;\sigma)$ is the return probability at $x$. For a finite manifold volume $V$ the average 
return probability $K_g(\sigma)$ can be written as
\beq\label{j5-6}
K_g(\sigma) = \frac{1}{V}
\int_M d^d x \sqrt{g(x)}\;K_g(x,x;\sigma) = \frac{1}{(4\pi \sigma)^{D_s/2}} \sum_{r=0}^\infty
A_r \sigma^r,\quad  A_0=1,
\eeq
where the coefficients $A_r$ can be expressed as integrals over local invariants,
like suitable contractions of powers of the curvature tensor. In eq.\ (\ref{j5-6}),
$D_s$ is the \textit{spectral dimension}, which for a smooth manifold $M$ satisfies
$D_s = d$. Diffusion processes can also be defined on more general, nonsmooth metric spaces 
by using an appropriate implementation of the diffusion equation (\ref{j5-5}), allowing us to
extract a spectral dimension. 

\begin{figure}[t]
\centerline{\scalebox{0.8}{\rotatebox{0}{\includegraphics{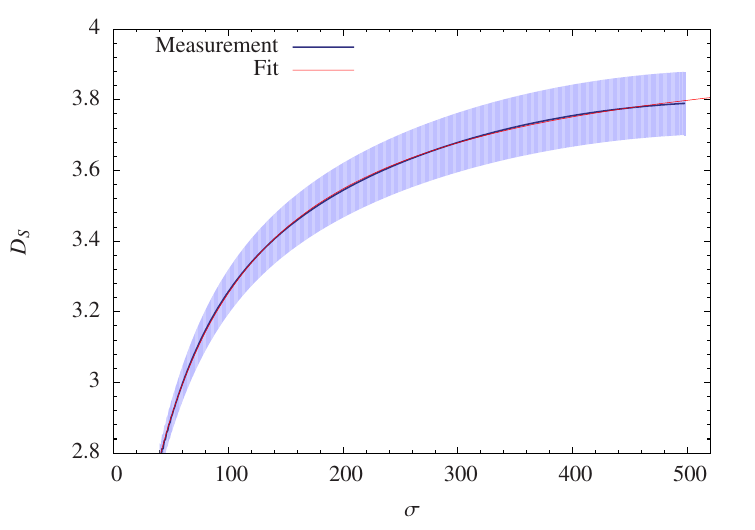}}}}
\caption{Spectral dimension $D_s$ as a
function of the diffusion time $\sigma$, measured for a spacetime volume
$N_4\! =\! 360.000$. The averaged measurement data lie along the central curve,
together with a superimposed best fit
$D_s(\sigma)\! =\! 4.02 - 119/(54+\sigma)$, and the two outer curves represent error bars.}
\label{figj4}
\end{figure}

Since $K_g(\sigma)$ is invariant under diffeomorphisms, it makes sense to consider its
quantum average $\langle K_g(\sigma)\rangle$ in a theory of quantum gravity, including on
the configurations of CDT lattice gravity. 
The expectation value $\la K_g(\sigma) \ra_{V_4}$ at constant volume $V_4$ has been measured in the
de Sitter phase of CDT, using Monte Carlo simulations \parencite{reconstruct,spectral}. The extracted spectral dimension
as a function of the diffusion time $\sigma$ is shown in Fig.\ \ref{figj4}.
The fact that $D_s$ depends nontrivially on $\sigma$ came as a surprise. 
Since some power of $\sigma$ will be a measure of the linear size of the region explored by the
diffusion process, small values of $\sigma$ correspond to short-distance excursions from the starting point $x$,
characterizing the short-scale dimension of quantum spacetime. Similarly, large diffusion times $\sigma$ are 
associated with long-distance properties.  

In flat space, only the first term on the right-hand side of \rf{j5-6} will contribute, as long as $\sigma$ is not too large,
such that the measurements are not subject to finite-size effects. If typical spacetime geometries were
fractal and self-similar up to some scale, 
one would still expect the first term in the expansion to dominate, leading to an initial plateau of the curve for $D_s$, possibly
at a value $D_s\not= 4$. Instead, we observe a nontrivial, monotonically increasing spectral dimension. 
Reassuringly, the asymptotic value for large $\sigma$, extrapolated from the measurements, is compatible with 4, the correct
classical limit. 

Equally important and even more intriguing is the observation of a genuine quantum signature for the short-scale dimension,
which extrapolated from the data is given by $D_s(\sigma\rightarrow 0)=1.80\pm 0.25$, a phenomenon dubbed 
\textit{dynamical dimensional reduction}. This value, which may be interpreted
as evidence that quantum spacetime is effectively two-dimensional in the ultraviolet (UV), is remarkable since it could indicate
a nonperturbative resolution of quantum gravity's nonrenormalizability. Since the initial discovery of this effect, there have been numerous
attempts to compute the spectral dimension in other approaches to quantum gravity, which have largely found a similar
behaviour. It suggests that the short-distance dimensional reduction could be a universal feature of quantum gravity, as advocated
by \textcite{carlipdim}. 

It is worth emphasizing that the spectral dimension for $\sigma\rightarrow 0$ is a rare instance of a 
nonperturbative gauge-invariant observable that is simply a number, which can be calculated and compared across
approaches. Although this result has not been related explicitly to phenomenology, it represents genuine progress in a field that 
used to be characterized by a complete absence of computable observables beyond perturbation theory. 

Another way to characterize a metric space $M$ is by its \textit{Hausdorff dimension} $d_H$, which we will take to mean the 
growth rate of the volume $V(B_r)$ of geodesic balls $B_r$ as a function of their radius $r$ and averaged over $M$, 
\begin{equation}
\bar{V}(B_r)\propto r^{d_H}.
\label{haus}
\end{equation} 
All of the required ingredients exist on the
triangulations we are interested in. This includes a notion of geodesic \textit{link distance} between lattice vertices $x$ and $y$, defined as the
number of one-dimensional links (lattice edges) in the shortest path of consecutive links connecting $x$ and $y$. The volume $V(B_r)$ is
then given by the number of vertices with link distance smaller or equal to $r$ from a given vertex. The expectation value of the Hausdorff
dimension in the de Sitter phase of CDT was measured by \textcite{reconstruct} on geometries with volumes of up to $N_4=180.000$ and found to be 
$d_H= 4.01\pm 0.05$ over a wide range of scales. We conclude that this particular observable exhibits classical behaviour throughout, in the
sense of expectation values.  

\subsection{Curvatures}
\label{sec:curv}

The central role of curvature in classical general relativity raises the question of whether something similar could be true in quantum gravity also.
Since the Riemannian curvature is a complicated rank-4 tensor, which depends on second derivatives of the metric and
encodes a large amount of local curvature data, one may immediately anticipate difficulties in representing it in the quantum theory.
The challenge beyond perturbation theory is to construct diffeomorphism-invariant curvature operators that are appropriately regularized and
renormalized. Recalling the nowhere-differentiable character of the quantum geometry in CDT, it is unclear a priori whether a meaningful
notion of curvature can be defined in a nonperturbative regime at all. As we will see below, this has now been realized. 

\begin{figure}[t]
\centerline{\scalebox{0.52}{\rotatebox{0}{\includegraphics{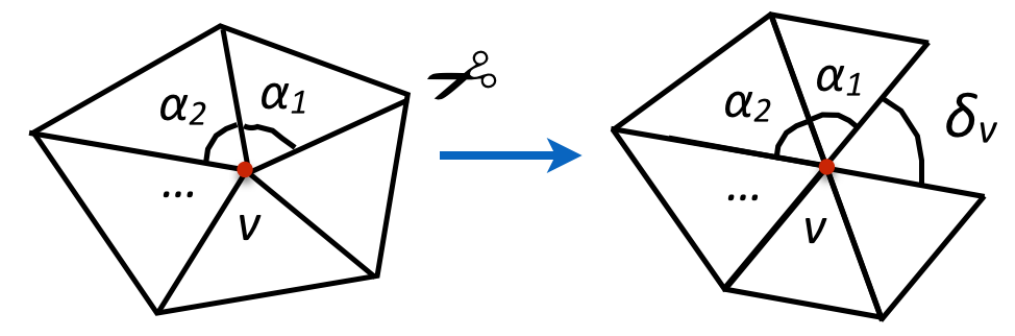}}}}
\caption{Each of the triangles meeting at the vertex $v$ contri\-butes an angle $\alpha_i$. Cutting the two-dimensional triangulation open as indicated and 
flattening it out on a straight surface reveals a deficit angle $\delta_v$.}
\label{fig:deficit}
\end{figure}

Before presenting this new notion of quantum curvature, let us review some well-known facts about finite piecewise flat manifolds 
and the singular curvature assignments they carry. 
A $d$-dimensional simplicial manifold, like the ones occurring in EDT or CDT, can be thought of as a nonsmooth metric space of Euclidean or Lorentzian
signature, whose individual triangular building blocks are flat by construction. This implies that curvature can only reside along lower-dimensional
(sub-)simplices, which turn out to be the $(d-2)$-dimensional simples, so-called \textit{hinges}, as already 
mentioned in Sec.\ \ref{sec:eubui}. The
presence of curvature at such a hinge is exhibited by parallel-transporting a vector around the hinge along a small closed loop $\gamma$ in a plane 
perpendicular to the hinge. This is most easily
visualized in a two-dimensional triangulation, where the hinges are the (zero-dimensional) vertices. The \textit{deficit angle} $\delta_v$ 
at a vertex $v$ is the difference between $2\pi$ and the sum of the angles of the triangles meeting at $v$ (Fig.\ \ref{fig:deficit}). It is a direct measure of the intrinsic,
Gaussian curvature located at $v$, and is also equal to the rotation angle that appears in the \textit{holonomy} of the Levi-Civita connection $\Gamma$ along $\gamma$,
\begin{equation}
U_\gamma(\Gamma):= \mathrm{P}\exp\Big( -\oint_\gamma \Gamma \Big), 
\label{holo}
\end{equation}
which describes how a tangent vector is rotated during its parallel transport along $\gamma$, where P denotes path ordering. 
The construction of holonomies (\ref{holo}) and their associated Wilson loops in CDT in $d\! =\! 4$ is described in \textcite{wilsonloop}.

\textcite{regge} has given a prescription for the total scalar curvature of a simplicial manifold $M$ in terms of
a weighted sum over all deficit angles in $M$. For an equilateral EDT configuration, this expression -- together with the cosmological-constant term --
gives rise to the form (\ref{j3-1}) of the Einstein-Hilbert action, and similarly in CDT for the Lorentzian version (\ref{j3-4a}) of the action. 
However, for reasons recapitulated in \textcite{curvreview}, it turns out that
the integrated scalar curvature \`a la Regge diverges in the continuum limit and is therefore not a good curvature observable in four-dimensional EDT 
or CDT quantum gravity.

An alternative to the deficit-angle curvature, which is better behaved in the UV, has been discovered recently. This so-called \textit{quantum Ricci
curvature} relies purely on local distance and volume measurements and not on the availability of tensor calculus. 
It can nevertheless also be applied to smooth Riemannian manifolds, where in the limit of infinitesimal distances it reproduces
the local Ricci curvature $Ric(v,v)\! =\! R_{\mu\nu}v^\mu v^\nu$ associated with a normal vector $v^\mu$ \parencite{qrc1,qrc2}. However, its key
strength is the applicability to a much larger class of nonsmooth metric spaces, including simplicial manifolds. 

\begin{figure}[t]
\centerline{\scalebox{0.5}{\rotatebox{0}{\includegraphics{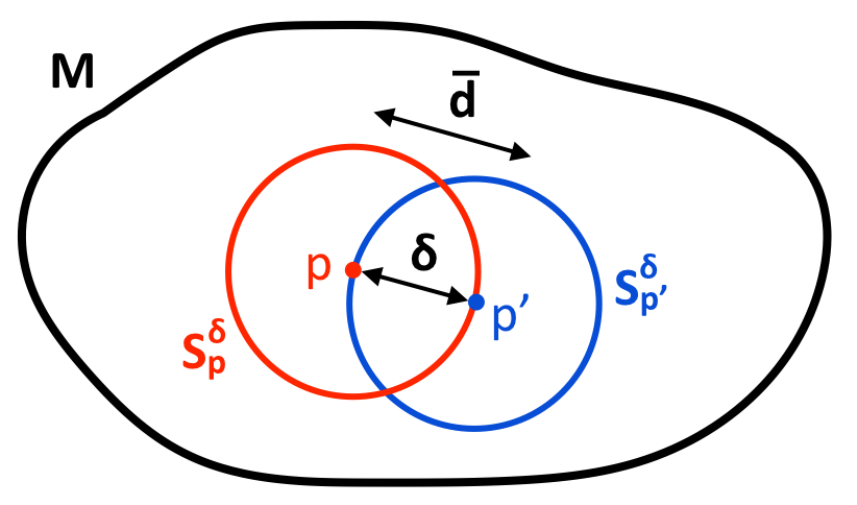}}}}
\caption{Comparing the distance $\bar{d}(S^\delta_p,S^\delta_{p'})$ of two $\delta$-spheres $S_p^\delta$ and
$S^\delta_{p'}$ with the distance $\delta$ of their centres $p$ and $p'$.}
\label{fig:riccispheresnew}
\end{figure}

The main ingredient in its construction is the \textit{average sphere distance} $\bar{d} (S^\delta_p,S^\delta_{p'})$ between 
two spheres of geodesic radius $\delta$, whose centres $p$ and $p'$ are also a distance $\delta$ apart.
On a smooth four-dimensional Riemannian manifold $(M,g_{\mu\nu})$ it is given by
\begin{equation}
\bar{d} (S^\delta_p,S^\delta_{p'}):=\frac{1}{vol(S_p^\delta)}\frac{1}{vol(S^\delta_{p'})}\int_{S^\delta_p}d^{3}\! q\,\sqrt{{\det h}}\int_{S^\delta_{p'}} 
d^{3}\! q'\sqrt{\det h'}\ d_g(q,q'),
\label{spheredist}
\end{equation}
where $d_g(q,q')$ denotes the geodesic distance of the points $q$ and $q'$,
and $h$ and $h'$ are the metrics induced on the two spheres (Fig.\ \ref{fig:riccispheresnew}). 
The quasi-local \textit{quantum Ricci curvature $K_q$ at scale} $\delta$
is then defined in terms of the quotient of the average sphere distance (\ref{spheredist}) and the centre distance as
\begin{equation}
\bar{d} (S_p,S_{p'})/\delta =c_q (1-K_q(p,p')), \;\;\; \delta=d_g(p,p'),
\label{ricdefine}
\end{equation}
where $c_q$ is a nonuniversal $\delta$-independent constant with $0\! <\! c_q\! <\! 3$, depending on the type and dimension of 
the metric space, and $K_q$ captures the nonconstant remainder. 

The simplest way of constructing a curvature observable for a given $M$ is by integrating the quotient (\ref{ricdefine}) over all pairs
$(p,p')$ of points at fixed distance $\delta$. Important for the implementation in quantum gravity is that the sphere distance (\ref{spheredist})
is equally well defined on CDT geometries. It allows for the evaluation of the expectation value $\langle \bar{d}(\delta)/\delta \rangle$
as a function of the geodesic length scale $\delta$, the so-called \textit{curvature profile}. The curvature profile in CDT quantum gravity in $d\! =\! 4$ 
in a near-Planckian regime has been measured and found to be compatible with that of a classical de Sitter space in \textcite{qrc3}. This confirms 
the viability of the quantum Ricci curvature in nonperturbative quantum gravity and the de Sitter-nature of the dynamically generated
quantum spacetime, both highly nontrivial results.

The availability of a notion of Ricci curvature which is well-defined in a nonperturbative regime and can in principle pick up direction-dependence  
(via the relative position of the two spheres) is significant. It opens the door to investigating new quantum observables that will lead to
a more fine-grained understanding of the properties of quantum spacetime. Examples are diffeomorphism-invariant measures of homogeneity
\parencite{homog} and correlation functions \parencite{conncorr}. This is important from a physical point of view since the
de Sitter-like quantum universe provides a model for the very early universe, and moreover has been derived from first principles. Its predictions should
be compared with the ad-hoc assumptions made in standard early-universe cosmology regarding the homogeneous and isotropic nature of
spacetime and the spectrum of the quantum fluctuations that act as seeds of structure formation. A re-examination of these assumptions 
from the perspective of nonperturbative quantum gravity may give us new insights into whether or not they are justified.

\vspace{1.0cm}
\noindent\textbf{Acknowledgement.} We are most grateful to the numerous collaborators and colleagues who contributed to the results
reported here and elsewhere. 

\printbibliography

\end{document}